\pdfoutput=1
\documentclass[aps,prd,amsmath,floats,floatfix, twocolumn,
superscriptaddress,nofootinbib,showpacs]{revtex4-1}

\usepackage[T1]{fontenc}
\usepackage[utf8]{inputenc}
\usepackage{lmodern}

\usepackage{verbatim}

\usepackage[dvipsnames, usenames]{xcolor}
\definecolor{linkcolor}{rgb}{0.0,0.3,0.5}
\usepackage[hypertexnames=false, unicode, colorlinks=true, linkcolor=linkcolor,
citecolor=linkcolor, filecolor=linkcolor,urlcolor=linkcolor,
pdfusetitle]{hyperref}

\usepackage[all]{hypcap}
\usepackage{graphicx}
\usepackage{xspace}
\usepackage{amssymb}
\usepackage[normalem]{ulem} 
\usepackage{bm} 
\usepackage{enumitem,amssymb}

\usepackage{microtype}

\usepackage[english]{babel}
\usepackage{blindtext}

\graphicspath{%
  {figs/}%
}

\DeclareMathAlphabet{\mathpzc}{OT1}{pzc}{m}{it}

\newlist{todolist}{itemize}{2}
\setlist[todolist]{label=$\square$}

\begin{document}

\title{Methodology for constraining ultralight vector bosons with gravitational wave searches targeting merger remnant black holes}

\author{Dana Jones}
\email{dana.jones@anu.edu.au}
\affiliation{OzGrav-ANU, Centre for Gravitational Astrophysics, College of Science, The Australian National University, Australian Capital Territory 2601, Australia}

\author{Nils Siemonsen}
\affiliation{Princeton Gravity Initiative, Princeton University, Princeton NJ 08544, USA}
\affiliation{Department of Physics, Princeton University, Princeton, NJ 08544, USA}

\author{Ling Sun}
\affiliation{OzGrav-ANU, Centre for Gravitational Astrophysics, College of Science, The Australian National University, Australian Capital Territory 2601, Australia}

\author{William E. East}
\affiliation{Perimeter Institute for Theoretical Physics, Waterloo, Ontario N2L 2Y5, Canada}

\author{Andrew L. Miller}
\affiliation{Nikhef -- National Institute for Subatomic Physics,
Science Park 105, 1098 XG Amsterdam, The Netherlands}
\affiliation{Institute for Gravitational and Subatomic Physics (GRASP),
Utrecht University, Princetonplein 1, 3584 CC Utrecht, The Netherlands}

\author{Karl Wette}
\affiliation{OzGrav-ANU, Centre for Gravitational Astrophysics, College of Science, The Australian National University, Australian Capital Territory 2601, Australia}

\author{Ornella J. Piccinni}
\affiliation{OzGrav-ANU, Centre for Gravitational Astrophysics, College of Science, The Australian National University, Australian Capital Territory 2601, Australia}

\hypersetup{pdfauthor={Jones et al.}}

\date{\today}

\begin{abstract}
Ultralight bosons are a hypothetical class of particles predicted under various extensions of Standard Model physics. As a result of the superradiance mechanism, we expect ultralight bosons, should they exist in certain mass ranges, to form macroscopic clouds around rotating black holes, so that we can probe their existence by looking for the long-transient gravitational wave emission produced by such clouds. In this paper, we propose a statistically robust framework for constraining the existence of ultralight vector bosons in the absence of detecting such a signal from searches targeting merger remnant black holes, effectively marginalizing over the uncertainties present in the properties of the target black holes. We also determine the impact of weak kinetic mixing with the ordinary photon and vector mass generation through a hidden Higgs mechanism on the constraining power of these searches. 
We find that individual follow-up searches, particularly with the next-generation gravitational wave detectors, can probe regions of parameter space for such models where robust constraints are still lacking.
\end{abstract}
\maketitle

\section{Introduction}
\label{sec:introduction}

Ultralight bosons are a hypothetical class of particles beyond the Standard Model that have been invoked as solutions to open questions in particle physics and cosmology. Current observation-oriented research is mainly centered on two sub-classes of ultralight boson: scalar (spin-0) bosons, such as the QCD axion, which provide well-motivated solutions to the strong CP-problem~\cite{Peccei_Quinn1977, Peccei_Quinn1977_2, Weinberg1978}, or other axion-like particles~\cite{Arvanitaki2010}; and vector (spin-1) bosons, which, for instance, arise in low-energy limits of quantum gravity models~\cite{Goodsell2009, Holdom1986, Jaeckel2010, Essig2013, Hui2017, Agrawal2020, Fabbrichesi2021, Bustillo2021, Bustillo2023}. To be able to detect these conjectured particles in a traditional lab setting, one must assume scenarios with certain couplings, however weak, to the Standard Model. However, with the phenomenon known as black hole (BH) superradiance, one can probe the existence of ultralight bosons by simply assuming a minimal coupling to gravity.

Through the superradiance mechanism, ultralight bosons can form gravitationally bound field configurations around a spinning BH, growing to form a macroscopic cloud as mass-energy is extracted from the BH~\cite{Penrose1969, Press_Teukolsky1972, Zel'Dovich1971, Starobinskii1973, Detweiler1980, Bekenstein1973, Dolan2007, Arvanitaki2011}. The superradiant mechanism of a bosonic field configuration of frequency $\omega$ and azimuthal number $m$ operates efficiently when two conditions are met: a) The BH's horizon frequency $\Omega_H$ is greater than that of the bosonic field configuration, $\omega < m \Omega_H$, and b) the Compton wavelength of the boson is comparable to the BH's horizon radius. The latter implies that the dimensionless coupling constant $\alpha=G M m_b$, which links the BH mass $M$ and boson mass $m_b$, satisfies $\alpha\lesssim\mathcal{O}(1)$. (Here and in the following, we employ $c=1$ and $\hbar=1$ units.) Systems meeting these criteria undergo a phase of exponential growth of the boson cloud, saturating gravitationally by spinning down the BH until condition a) is roughly saturated.
What remains is a boson cloud with a mass of at most 10\% of the BH's mass, which dissipates as it emits gravitational waves (GWs)~\cite{Arvanitaki2011, Yoshino2014, Yoshino2015, Arvanitaki2015, Arvanitaki2017, Brito2017, Brito2017_2, Yoshino2014, Yoshino2015, Baryakhtar2017, Chan2022, Cardoso2018, Hannuksela2019, Zhang2019, East2017, East2017_2, East2018, Siemonsen2020,Brito:2015}. This radiation is quasi-monochromatic with a frequency of roughly twice the oscillation frequency of the boson cloud~\cite{Arvanitaki2015} and a slight upward frequency drift~\cite{Baryakhtar2017}. While the superradiant instability is in principle possible for a BH of any mass, granted a boson particle of appropriate mass exists, our interest lies with stellar-mass BHs, whose matching bosons emit at frequencies falling within the sensitive band of both existing ground-based GW detectors (i.e., the Advanced Laser Interferometer Gravitational-Wave Observatory (aLIGO)~\cite{aLIGO}, Advanced Virgo~\cite{aVirgo}, and KAGRA~\cite{KAGRA}) and future ones (e.g., Einstein Telescope~\cite{Punturo2010, Hild2011, ET-design, Maggiore2020} and Cosmic Explorer~\cite{Abbott-CE, Evans2021, Reitze2019}). In targeting stellar-mass BHs, one can probe a boson mass range from roughly $10^{-14}$~eV to $10^{-11}$~eV~\cite{Arvanitaki2015, Brito2017_2}.

To date, several searches and studies have been carried out to either uncover or rule out the existence of scalar and/or vector boson clouds. BH spin measurements are used to constrain the existence of the ultralight scalars~\cite{Arvanitaki2015, Ng2021, Cardoso2018, Brito2017_2} and vectors~\cite{Baryakhtar2017, Cardoso2018}, but significant uncertainties are associated with these studies.
Blind searches targeting the whole sky or galactic center~\cite{O3_all-sky_scalar_bosons, Palomba2019, Dergachev2019,O3_CWs_Milky_Way, Zhu2020} place constraints on the existence of scalars, roughly disfavoring the mass range of $\sim 10^{-13}$--$10^{-12}$~eV, under certain astrophysical assumptions.
A directed search targeting the BH in the X-ray binary Cygnus X-1 disfavors scalars in the mass range $\sim 0.6$--$1\times 10^{-12}$~eV, depending on the assumed age of the BH~\cite{O2_CygnusX1_scalar_bosons, Collaviti2024}.
Studies have used the null results of searches for a stochastic GW background from a population of BHs hosting scalar~\cite{Tsukada2019} and vector~\cite{Tsukada2021} boson clouds, resulting in disfavored mass ranges of $2.0$--$ 3.8 \times 10^{-13}$~eV and $0.8$--$6.0 \times 10^{-13}$~eV for scalars and vectors, respectively.
However, all of these studies make certain assumptions about the underlying BH population, including the unknown age and spin of the target BHs, so the derived constraints should be interpreted with caution.
On the other hand, constraints from boson searches targeting remnant BHs formed in binary mergers that are observed in GWs are more robust, as there exists more complete knowledge of the BH parameters and history. We focus on merger remnant BHs (see, e.g., Refs.~\cite{GWTC-1, GWTC-2, GWTC-3}) in this paper.

Reference~\cite{Isi2019} shows that one can use continuous wave (CW) analysis techniques, which were initially designed to search for gravitational waves from spinning non-axisymmetric neutron stars, to search for signals from scalar clouds around merger remnant BHs. However, these targets generally lie outside the sensitivity range of current-generation detectors. 
In Ref.~\cite{Jones2023}, it is demonstrated that vector clouds around merger remnant BHs, which emit stronger but shorter GW signals compared to scalar clouds, can potentially be reached with current-generation detectors using techniques adapted from conventional CW searches.
The directed search method in Ref.~\cite{Jones2023} for targeting vector signals from merger remnant BHs is adapted from the semicoherent hidden Markov model (HMM) search method used in more traditional CW searches (see, e.g., Refs.~\cite{Suvorova2016, Sun2018, Isi2019, Sun2019}).
Reference~\cite{Jones2023} also assumes that the parameters of the target BH (the mass, spin, sky position, and orientation) are perfectly known, so that for a given boson mass, one can determine the expected GW signal with a sufficiently good model~\cite{Siemonsen2022_2}. In this case, constraining the boson mass in the absence of a detection is fairly straightforward. However, in reality, no system parameter is perfectly known; the best knowledge of the remnant BH comes from the posterior distribution obtained through compact object coalescence (CBC) analysis of the merger event~\cite{first_detection, Veitch2015}. In this paper, we formulate a procedure to incorporate these parameter uncertainties when inferring constraints on the boson mass in the absence of a detection.

These search techniques assume only a minimal coupling of the vector boson to gravity. However, as a vector boson cloud grows to large field values through superradiance, couplings between the vector field and the Standard Model, or other nonlinear interactions besides gravity, may become important in certain particle models. For instance, sufficiently strong self-interactions may halt the scalar superradiance process and invalidate constraints based on null observations \cite{Arvanitaki2011, Fukuda2019, Baryakhtar2020, Yoshino2014, Yoshino2015, Omiya2022, Collaviti2024}. In the case of vector bosons, non-vanishing kinetic mixing with the Standard Model photon can lead to the production of a pair plasma and subsequent radiation of electromagnetic emissions \cite{Siemonsen2022} (see also Ref.~\cite{Xin2024}). If the vector's mass is obtained through a dark Higgs mechanism, then an additional dissipation channel opens up \cite{Fukuda2019} or a stringy bosenova ensues \cite{East2022,East2022_2}. In both cases, the additional interactions can significantly impact the superradiance process, GW emission, and hence, constraints obtained from superradiance signatures. Therefore, here we relax the assumptions made in standard vector GW searches---that is, that these effects are negligible---and discuss how to extend the constraints to kinetically mixed dark photons and a dark Higgs-Abelian sector. Finally, we forecast the parameter ranges of these two extended dark sectors accessible by CBC remnant follow-up searches using next-generation ground-based detectors.

The layout of the paper is as follows. In Sec.~\ref{sec:method}, we review the search method for vector signals and describe the procedure of deriving constraints in the absence of a detection when the BH parameters are uncertain. In Sec.~\ref{sec:simulations}, we address the practical aspects of the method in detail using illustrative simulations. In Sec.~\ref{sec:other_models}, we discuss alternative interpretations of the constraints in two particular models for dark photon interactions, a kinetic mixing and a dark Higgs mechanism. 
We summarize our results and discuss future implications in Sec.~\ref{sec:conclusion}.


\section{Inference method}
\label{sec:method}

\subsection{Long-transient search method}
\label{sec:search_method}

In Ref.~\cite{Jones2023}, we propose a search method specially designed to track long-transient signals produced by ultralight vectors. This method employs an HMM tracking scheme, as do many contemporary CW search methods, but it is able to track signals with frequency evolution timescales of $\sim \mathcal{O}$(minutes--days), which are considerably shorter than those seen in a typical CW search. To generate synthetic signal waveforms, we use the waveform model \texttt{SuperRad}, described in Ref.~\cite{Siemonsen2022_2}. We demonstrate that, in a (nearly) optimal scenario, a vector boson signal generated by a system with BH mass $M \gtrsim 60 M_{\odot}$, dimensionless spin $\chi \gtrsim 0.6$, and distance $d \lesssim 1$~Gpc should lie within the detection horizons of current-generation detectors~\cite{Jones2023}.

The search method~\cite{Jones2023} is summarized as follows. We take the short Fourier transforms (SFTs) of the time series data from a given GW detector over the relevant stretch of observing time $T_{\rm obs}$ during which we expect a signal may be detectable. We use these SFTs to compute a frequency-domain matched filter ($\mathcal{F}$-statistic) over a coherent time interval $T_{\rm coh}$, set by the frequency drift of the signal. Then, we combine these segments of length $T_{\rm coh}$ incoherently with the HMM-based tracking algorithm. We define the detection statistic as the total log-likelihood of the optimal path returned by the HMM algorithm, divided by the number of coherent segments in the search, i.e., Eq.~(24) in Ref.~\cite{Jones2023}. Candidates with a detection statistic above the threshold for a chosen false alarm probability, $P_{\rm fa}$, are identified and followed up using a similar kind of validation process to what has been done in previous CW searches (see, e.g., Refs.~\cite{Sco_X-1_O3, O3_young_supernova_remnants}).

For a given BH-boson system (with vector mass $m_V$), we choose the optimal search configuration, detailed in Ref.~\cite{Jones2023}.
As the BH mass increases, the optimally matched boson mass $m_V^{\rm opt}$\footnote{For a particular BH with a given mass and spin, the optimally matched boson mass $m_V^{\rm opt}$ (with corresponding dimensionless coupling constant $\alpha_{\rm opt}$) maximizes the superradiant instability, yielding the largest possible GW strain amplitude when the cloud reaches its maximum size.}
decreases, as does the GW signal frequency.
Since the search sensitivity scales with $T_{\rm coh}$~\cite{Sun2018}, for a given BH-boson system with a maximum estimated spin-up $\dot{f}^{\rm max}_{\rm GW}$, we choose the largest possible $T_{\rm coh}$ which satisfies the following relation:
\begin{equation}
    T_{\rm coh} \leq (2 \dot{f}^{\rm max}_{\rm GW})^{-1/2},
    \label{eqn:Tcoh}
\end{equation}
such that the signal does not drift outside the frequency bin, $\Delta f = 1/(2T_{\rm coh})$, over each coherent time segment.
We set the following bounds on the SFT length $T_{\rm SFT}$ and coherent length $T_{\rm coh}$: $0.25~{\rm min} \leq T_{\rm SFT} \leq 30~{\rm min}$, and $1~{\rm min} \leq T_{\rm coh} \leq 10~{\rm day}$. (See Sec.~IIIA in Ref.~\cite{Jones2023} for the motivations behind these bounds.) 
Moreover, we require that a single detector contributes at least four SFTs to a single coherent segment, i.e., $4\,T_{\rm SFT} \leq T_{\rm coh}$.

\subsection{CBC parameter estimation}
\label{sec:cbc_pe}
This paper builds upon Ref.~\cite{Jones2023}, focusing on the case in which we target a CBC merger remnant BH with parameters that have some baked-in uncertainties, leading to uncertainties in the boson signal model, search configuration, and results. 
In particular, we formulate a statistically robust procedure for deriving constraints on the existence of vectors in the absence of a detection.

We start by briefly summarizing how the BH parameter uncertainties that must be accounted for arise.
When one or more detectors identify a candidate signal, different types of searches are run to determine if it is a real GW signal or simply noise. One type of search attempts to recover the signal assuming the candidate was the result of a CBC~\cite{GW150914_method}. It uses optimal matched filtering with a template bank of waveforms predicted by general relativity.
The other type of search does not make any particular assumptions about the signal waveform; instead, it considers a variety of generic transient signals~\cite{Klimenko2008, Klimenko2016}. Although the methods used in these two types of searches are independent, we would expect sufficiently loud CBC signals to be recovered by both~\cite{first_detection}.

Once a CBC merger is detected, numerical models derived from general relativity are used to generate relatively accurate parameter estimates. A coherent Bayesian analysis is performed for each model, using stochastic sampling techniques such as the Markov chain Monte Carlo and nested sampling algorithms, in order to derive posterior distributions of the physical system parameters~\cite{Veitch2015, Ashton2024, RomeroShaw2020, Smith2020, Skilling2006, Speagle2020}. However, uncertainties still remain in the estimates for the system's parameters due to a combination of statistical uncertainties and systematics which come from averaging together the parameter estimates from multiple different waveform models~\cite{first_detection}. In a follow-up search targeting the CBC remnant, like the one in this study, where results hinge on assumptions made about the source parameters, these uncertainties must be taken into account.
In principle, the posterior distribution of CBC parameters also depends on the priors set in the parameter estimation procedure. However, the events considered relevant for follow-up in boson searches have relatively high signal-to-noise ratios (SNRs) and hence are not prior dominated.

\subsection{Constraining the boson mass}
\label{sec:constraints}

In a typical CW search, the upper limits on the signal strain (and the corresponding source properties) in the absence of a detection are usually derived in a frequentist way, as outlined in, e.g., Ref.~\cite{Sco_X-1_O3}, and summarized below.
We start by considering a sub-band frequency range (e.g., 1 Hz) within which the search is carried out, and we assume no signal has been detected. The target source is characterized by a multi-dimensional vector parameter $\boldsymbol{\theta} = \{{\theta}_1, {\theta}_2, \cdots, {\theta}_M \}$, where each component parameter is associated with a distribution. 
Then, we generate $N$ synthetic signals, each based on a randomly drawn set of system parameters $\boldsymbol{\theta}_i = \{{\theta_1}_i, {\theta_2}_i, \cdots, {\theta_M}_i \}$ with $i = 1, 2, \cdots, N$ and inject them into $N$ noise realizations. 
We repeat such injection study by setting the strain $h_0$ to different values in order to find the strain value $h_0^{\rm CL}$, with the superscript $\rm CL$ indicating the ratio of the injections recovered (for example, when ${\rm CL} = 95\%$, the value of $h_0^{95 \%}$ indicates the ``95\% confidence upper limit'' derived in a given sub-band).

To constrain the existence of ultralight boson particles in the case of a non-detection, we use a similar frequentist approach by incorporating the CBC parameter estimation results, as detailed below.
For a given remnant BH, we define the probability of detecting a signal generated by the BH-boson system with boson mass $m_V$ as
\begin{equation}
    P_{\rm det}(m_V) = \int P_{\rm det}(\boldsymbol{\theta}; m_V) ~ P(\boldsymbol{\theta}) \, d\boldsymbol{\theta} \leq 1,
    \label{eqn:P_continuous}
\end{equation}
where $P(\boldsymbol{\theta})$ is the posterior probability of a set of sample BH parameters $\boldsymbol{\theta}$, and $P_{\rm det}(\boldsymbol{\theta}; m_V)$ is the probability the signal is detectable given that it is generated by the BH with parameters \boldsymbol{$\theta$}, and given the existence of vector bosons with mass $m_V$. 
In other words, we calculate the probability of detecting a signal if vectors with a mass of $m_V$ exist, marginalized over the remnant BH parameters.

As described in Sec.~\ref{sec:cbc_pe}, the BH parameter distributions are available in the form of a set of multi-dimensional posterior samples, so in reality we calculate $P_{\rm det}(m_V)$ in a discrete manner. We rewrite Eq.~\eqref{eqn:P_continuous} in its discrete form,
\begin{equation}
    P_{\rm det}(m_V) = \sum_{i=1}^{N_{\rm BH}} P_{\rm det}(\boldsymbol{\theta}_i; m_V) ~ P(\boldsymbol{\theta}_i),
    \label{eqn:P_discrete}
\end{equation}
where $N_{\rm BH}$ is the total number of sampled BHs. Because each sampled BH with the set of parameters $\boldsymbol{\theta}_i$ is drawn with equal probability from the posterior distribution obtained in CBC parameter estimation (we simply get more samples with $\boldsymbol{\theta}_i$ closer to the multi-dimensional maximum a posteriori values), Eq.~\eqref{eqn:P_discrete} becomes
\begin{equation}
    P_{\rm det}(m_V) = \frac{1}{N_{\rm BH}} \sum_{i=1}^{N_{\rm BH}} P_{\rm det}(\boldsymbol{\theta}_i; m_V).
    \label{eqn:P_det_original}
\end{equation}
For a given BH with a set of parameters $\boldsymbol{\theta}_i$ and a boson mass $m_V$, the signal waveform is known. The detection probability $P_{\rm det}(\boldsymbol{\theta}_i; m_V)$ is equivalent to the recovery rate in $N_{\rm noise}$ noise realizations, i.e., 
\begin{equation}
    P_{\rm det}(m_V) = \frac{1}{N_{\rm BH} N_{\rm noise} } \sum_{i=1}^{N_{\rm BH}} N_{\rm det}(\boldsymbol{\theta}_i; m_V),
    \label{eqn:P_det}
\end{equation}
where $N_{\rm det}(\boldsymbol{\theta}_i; m_V)$ is the number of recovered signals in $N_{\rm noise}$ noise realizations for a set of given BH parameters and can take on values between $[0, N_{\rm noise}]$.
In the absence of a detection in the search, $P_{\rm det}(m_V)$ corresponds to the confidence level with which one can exclude the existence of vectors with mass $m_V$. We describe this procedure through detailed simulations in the following section.


\section{Simulations}
\label{sec:simulations}

In this section, we present the results of simulations to aid in illustrating the methods laid out in Sec.~\ref{sec:method}.
All synthetic waveforms are generated using \texttt{SuperRad}~\cite{Siemonsen2022_2}.
Assuming we do not have a detection in a search targeting a specific remnant BH formed in a merger event, we derive statistically robust constraints on the vector's existence as follows. We first discuss the required number of samples from the CBC posteriors to represent the distributions of the BH properties (Sec.~\ref{sec:convergence}). We then demonstrate the procedure of deriving constraints using a detailed synthetic target (Sec.~\ref{sec:synthetic_sys}).

\subsection{Number of BH samples}
\label{sec:convergence}

In a typical set of parameter estimation results of a CBC event, e.g., from the LIGO-Virgo-KAGRA catalogs~\cite{GWTC-1, GWTC-2, GWTC-3}, there are $\mathcal{O}(10^4)$ posterior samples which form the full multi-dimensional posterior distribution.
If we take all the posterior samples of the remnant BH, this yields $N_{\rm BH} \sim 10^4$ in Eq.~\eqref{eqn:P_det}.
For each set of sampled system parameters $\boldsymbol{\theta}_i$, corresponding to a particular definitive BH target, we usually quantify the recovery rate by simulating the system and injecting the signal into multiple noise realizations (e.g., $N_{\rm noise} \sim 10^2$), as described in Sec.~\ref{sec:constraints}. 
To derive the constraints on $m_V$, the simulation procedure should be repeated, calculating $P_{\rm det}(m_V)$ over a range of $m_V$ values in the most accessible parameter space, corresponding to $\sim 10$ values of $m_V$. 
Thus, in the worst case, $P_{\rm det}$ needs to be computed using a frequentist approach with $\mathcal{O}(10^7)$ trials to include both BH-boson system configurations and noise realizations, which is computationally infeasible. 
Instead, we randomly sample $N$ sets of parameters from the BH posterior distribution and quantify exactly how large $N$ needs to be to such that the subset of samples represents the full distribution reasonably well. 

Here, and in Sec.~\ref{sec:synthetic_sys}, we consider a synthetic system as an example, with parameters similar to GW170814~\cite{GW170814_detection}, an event in the second LIGO-Virgo observing run.
(However, the results and methodology generally apply to other CBC remnants.)
We first take the full posterior distribution of the remnant BH parameters from the GW170814 results~\cite{GWTC-1, GW170814_online, opendata_run12}.
Because this event is beyond the reach of vector searches with the sensitivity of the current detectors, we bring it closer, applying an offset of $-250$~Mpc to the luminosity distance $d$ of the real event (bringing the median of $d$ from 606~Mpc to 356~Mpc).
We also use the improved frequency-dependent detector noise amplitude spectral density (ASD) estimates from the third, rather than second, observing run~\cite{H1_asd, L1_asd}. We undertake these measures so that we can continue to use this example to demonstrate the procedure of constraining vector bosons.
The median, 5th percentile, and 95th percentile values for an array of the BH parameters are shown in Table~\ref{tab:GW170814}. 

\begin{figure*}[hbt!]
    \centering
    \includegraphics[scale=.5]{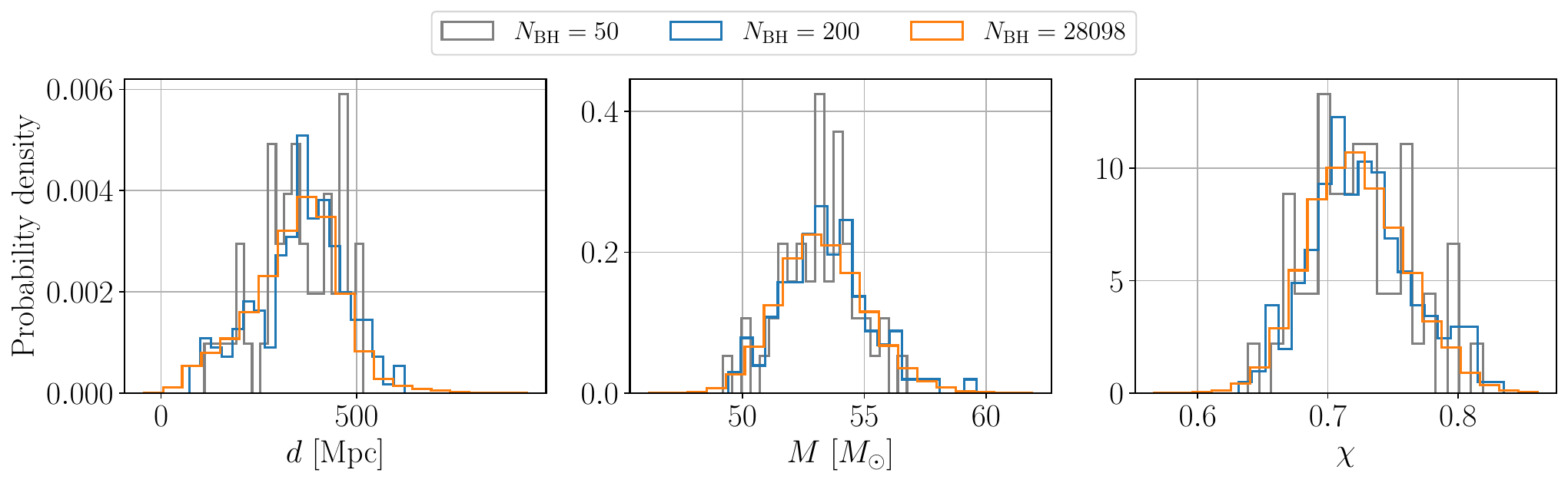}
    \caption{Probability density distributions for $d$, $M$, and $\chi$ of a GW170814-like system. The luminosity distance $d$ has been shifted 250~Mpc closer compared to the real event. The distributions of $M$ and $\chi$ are identical to the posterior distribution of GW170814. The orange curves show the full posterior distributions for each parameter, whereas the gray (blue) curves have been constructed from 50 (200) random samples drawn from the full distributions.}
    \label{fig:num_samples}
\end{figure*}

The distributions of subsets with $N$ samples are shown in Fig.~\ref{fig:num_samples} for three major BH parameters: $d$, $M$, and $\chi$.
With $N_{\rm BH} = 50$ samples (gray), the subset is not large enough to represent the full distribution (orange) well. The subset of $N_{\rm BH} = 200$ samples, however, is deemed large enough to represent the full distribution (i.e., the percent error of the variance for each $N_{\rm BH} = 200$ sample set is $\lesssim 10\%$ when compared to the variances of the full distributions) while also ensuring these studies remain computationally efficient.
In addition, although in conventional frequentist simulations $N_{\rm noise} \gtrsim 10^2$ may be needed to quantify the recovery rate in noise realizations, in this study, $P_{\rm det}$ (for a given $m_V$ value) is evaluated over $N_{\rm BH}N_{\rm noise}$ trials [see Eq.~\eqref{eqn:P_det}]. By taking into consideration uncertainties from both BH parameters and noise realizations, a total number of $N_{\rm BH}N_{\rm noise} \sim 10^3$ ensures that enough trials are used to obtain $P_{\rm det}$. 
Thus, we set $N_{\rm BH} = 200$ and $N_{\rm noise} = 10$; in this case, the uncertainties from BH parameters dominate.

\begin{table*}[tbh]
    \centering
    \setlength{\tabcolsep}{20pt}
    \renewcommand\arraystretch{1.2}
    \begin{tabular}{lcccccc}
        \hline
    Percentile & $M~[M_{\odot}]$ & $\chi$ & $d$~[Mpc] & $\cos \iota$ & RA [rad] & Dec [rad] \\
        \hline
        5th & 50.5 & 0.67 & 126 & $-0.86$ & 0.68 & $-0.88$ \\
        Median & 53.2 & 0.72 & 356 & 0.76 & 0.81 & $-0.79$ \\
        95th & 56.4 & 0.79 & 514 & 0.98 & 0.87 & $-0.61$ \\
        \hline
    \end{tabular}
    \caption{Remnant BH parameters for a GW170814-like event~\cite{GW170814_detection}. The luminosity distance $d$ has been shifted 250~Mpc closer compared to the real event. The distributions of all other parameters are identical to the posterior distributions of GW170814 from the GWOSC catalog~\cite{GW170814_online}.}
    \label{tab:GW170814}
\end{table*}

\subsection{Synthetic system and constraints}
\label{sec:synthetic_sys}

In this section, we detail the procedure of setting constraints on a range of $m_V$ using the synthetic system introduced in Sec.~\ref{sec:convergence}, and assuming a search targeting such a remnant BH does not yield a detection.

For a given CBC remnant BH, let us first determine a range of $m_V$ to be studied. 
Even with a single set of BH parameters, a range of boson masses can be tested. This is because bosons with masses close to the value that optimally matches the sample BH can yield GW emission at a nearly maximized strength and have equally good observational prospects (e.g., see Fig.~11 in Ref.~\cite{Jones2023}). 
However, as discussed above, the BH parameters also have associated uncertainties which must be considered.
In general, one can take $m_V \approx [0.6 , 1.1]m_V^{\rm opt}$ as the promising boson mass range, where $m_V^{\rm opt}$ here is the boson mass which maximizes the strain amplitude for the median values of the BH parameters.
In principle, for each sampled set of BH parameters, a different optimally matched boson mass can be obtained. 
The above vector mass range has been chosen empirically based on the estimated signal strength produced by bosons with different masses, which is also validated by Fig.~11 of Ref.~\cite{Jones2023}, in which the bright region corresponds to the boson mass range that produces potentially detectable signals for a given BH.
For BHs with masses within two standard deviations of the mean of the mass distribution, all corresponding optimally matched bosons should fall within this range. 
For the synthetic system (with median parameters $M=53.2\ M_{\odot}$ and $\chi = 0.72$), we have $m_V^{\rm opt} = 4.612 \times 10^{-13}$~eV, so the promising search range is $m_V = [2.767, 5.073] \times 10^{-13}$~eV. We take a sample of 11 evenly-spaced values from across this range,
$m_V = \{2.767, 2.998, 3.228, 3.459, 3.689, 3.920, 4.151, 4.381, 4.612, \\4.842, 5.073\} \times 10^{-13}$~eV, and we calculate the detection probability $P_{\rm det}(m_V)$ for each $m_{Vj}$ (where the subscript $j$ indexes the $m_V$ value in the array) according to the guidelines that follow.

To avoid bias caused by any assumptions made about the target BH's true parameters, 
for a given $m_V$ value, we randomly draw $N_{\rm BH} = 200$ sample parameter sets from the distribution of the GW170814-like remnant BH (Table~\ref{tab:GW170814}). We generate a synthetic signal based on each sample set of system parameters \{$\boldsymbol{\theta}_i$; $m_{Vj}$\}, and then we inject each signal into $N_{\rm noise} = 10$ independent Gaussian noise realizations using the simulateCW Python module in the LALPulsar library of LALSuite~\cite{lalsuite, swiglal}. This injection procedure is repeated for each of the $N_{\rm BH} = 200$ samples with parameters $\boldsymbol{\theta}_i$. 
When attempting to recover each injection, one must use the same search configurations that would have been used in the search run on real detector data. 
In practice, this looks like a grid of configuration setups (search start time $t_{\rm start}$, coherent time $T_{\rm coh}$, and total observing time $T_{\rm obs}$) chosen based on the interesting parameter space. 
For details on how to choose the configurations to adequately cover the parameter space, see Appendix~\ref{appendix:configs}.
In this example, we have 11 configurations. 
For any synthetic signal injection, it is considered a successful recovery as long as the injection is recovered by at least one configuration.
Using all 11 configurations, we calculate the recovery rate \mbox{$P_{\rm det}(m_{V_j})$} for a given $m_{V_j}$, out of $N_{\rm BH}N_{\rm noise} = 2000$ injections [Eq.~\eqref{eqn:P_det}]. [Note, due to the precision of \mbox{$P_{\rm det}(m_{V_j})$}, further increasing either the number of samples $N_{\rm BH}$ or the number of noise realizations $N_{\rm noise}$ is not beneficial.]
We repeat this process for each value of $m_{V_j}$.

\begin{figure}[hbt!]
    \centering
    \includegraphics[scale=.42]{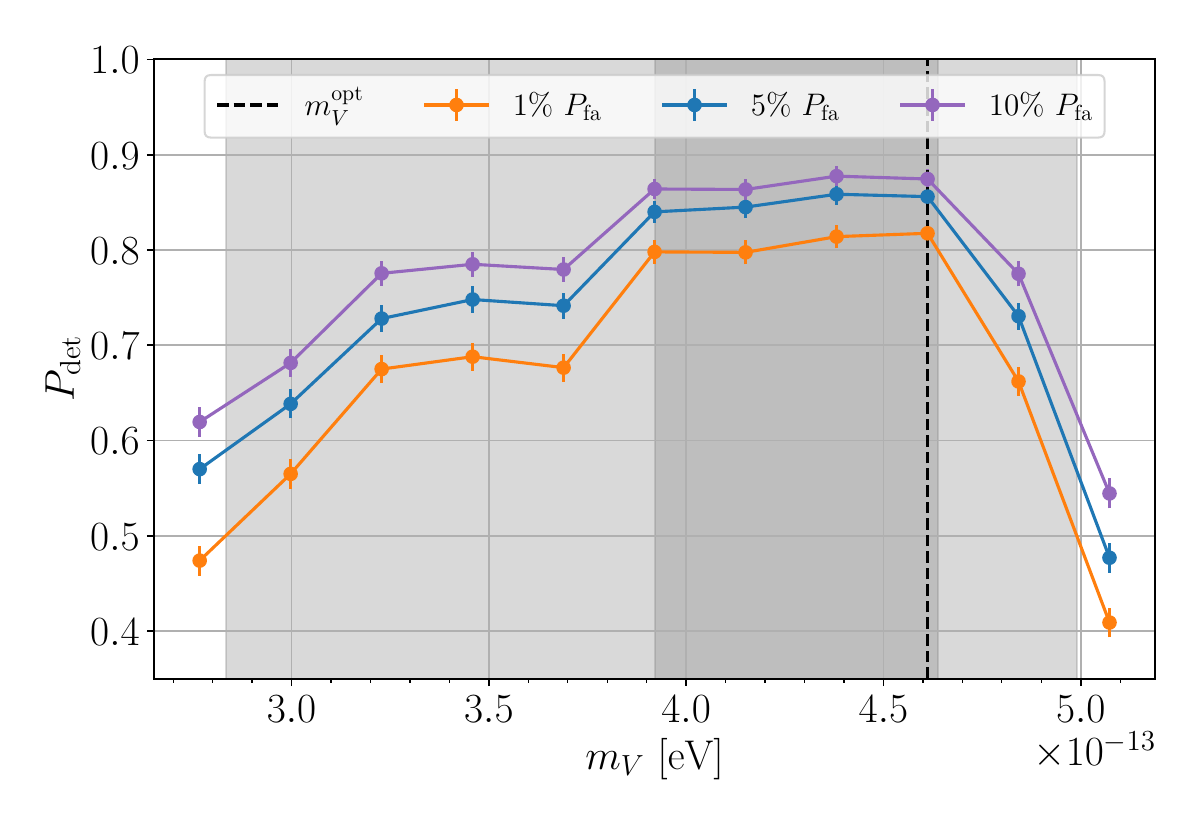}
    \caption{The detection probability $P_{\rm det}$ as a function of $m_V$ with a 1\% (orange), 5\% (blue), and 10\% (purple) false alarm probability. The vertical dashed line marks the optimal boson mass $m_V^{\rm opt}$ for the GW170814-like synthetic system with parameters shown in the median row in Table~\ref{tab:GW170814}. The dark gray (light gray) shaded region indicates the disfavored boson mass range with 80\% (50\%) confidence ($P_{\rm fa}=1\%$), assuming a search targeting such a remnant BH does not yield a detection. The error bars indicate the $1\sigma$ beta-binomial uncertainty in $P_{\rm det}$~\cite{beta-binomial}.}
    \label{fig:Pdet_v_mV}
\end{figure}

The results are displayed in Fig.~\ref{fig:Pdet_v_mV}, where $P_{\rm det}$ is shown as a function of $m_V$.
We note that all results shown here are for the synthetic event, and thus no real constraints on vector bosons are derived in this study.
Assuming that no detection is made from a search like this, a vector mass range of $[3.92, 4.64]$ ($[2.83, 4.99]) \times 10^{-13}$~eV can be excluded with 80\% (50\%) confidence, as indicated by the dark (light) gray shaded region.
The above simulated constraints are based on a 1\% $P_{\rm fa}$ threshold.
With a more relaxed false alarm probability (i.e., a higher $P_{\rm fa}$ threshold), the excluded regions become wider; Figure~\ref{fig:Pdet_v_mV} also shows the constraints corresponding to different false alarm probabilities, with the orange, blue, and purple curves indicating $P_{\rm fa} = 1\%$, 5\%, and 10\%, respectively. The constraints are more conservative for lower $P_{\rm fa}$ values.
 
One might notice that the curves in Fig.~\ref{fig:Pdet_v_mV} all plateau across certain boson mass ranges. These plateaus are the result of a balancing act that occurs between the intrinsic strain amplitude of the signal, which decreases as the boson mass is shifted to lower values, and the search sensitivity, which improves for lower boson masses that produce signals with smaller frequency drift and thus allow larger $T_{\rm coh}$ values to be used (also described in Ref.~\cite{Jones2023}). One can see a similar behavior in Fig. 11 of Ref.~\cite{Jones2023}, where the horizon distance peaks gradually over a range of vector masses rather than sharply over a single mass.

\section{Weakly coupled vector bosons}
\label{sec:other_models}

The standard assumption in interpreting search results for signatures of BH
superradiance and the presence of vector boson clouds is that the underlying
massive vector field only acts gravitationally and has no couplings to the
Standard Model or additional interactions. (In particular, one assumes that the
vector mass arises through the Stueckelberg
mechanism~\cite{stueckelberg1933,Ruegg2003}, as opposed to a Higgs mechanism.)
The inference method and the simulation example discussed in earlier sections
by default make this standard assumption.  In this section, we outline how this
assumption can be partially lifted, and how the constraints obtained from the
approach described above can be mapped to the regime where additional
interactions exist, but are not strong enough to nullify such searches.
Specifically, we focus on a kinetically mixed dark photon and a dark Higgs-Abelian sector as two well-motivated examples of vector interactions.

There are several ways that bosonic interactions can affect the evolution, and hence the GW signal, of a superradiant cloud:

\begin{enumerate}[label=\roman*.]
\item The interactions can change how the energy, and hence oscillation frequency, of the boson cloud depends on the number of bosons in the cloud. 
This will change the frequency of the GW signal.

\item The interactions can cause the cloud to saturate at a lower total energy than it would if the superradiant instability were to saturate gravitationally
(by spinning down the BH). This reduces the maximum amplitude of the resulting GW signal. This can happen due to the fact
that bosonic interactions introduce additional dissipation channels, e.g., due to bosonic radiation, that eventually cancel out the superradiant extraction
of energy through the BH.

\item Even if the extra dissipation channels due to the bosonic interactions are not strong enough to change the saturation of the superradiant instability,
they can still affect how quickly the cloud dissipates away after saturation. If the bosonic dissipation is stronger or comparable to gravitational radiation, it will affect how long the GW signal lasts and how quickly the frequency evolves.
\end{enumerate}

For the kinetically mixed dark photon (Sec.~\ref{sec:kineticmix}), we primarily consider when interactions with an electron-positron plasma cause the cloud to dissipate faster
(effect iii). For the Higgs-Abelian model (Sec.~\ref{sec:higgs}), all three of these effects are relevant to the GW signal. We consider how interactions mediated by the Higgs-like field can change the self-energy of the cloud and lead to bosonic radiation, as well as how, in some parts of the parameter space, the growth of the cloud will be halted due to a stringy bosenova.

In Sec.~\ref{sec:popproj}, we forecast the parameter space accessible to
next-generation GW detectors such as the Einstein Telescope and
Cosmic Explorer.

\subsection{Kinetically mixed dark photon}
\label{sec:kineticmix}

We focus first on a kinetic mixing $\varepsilon$ between the Stueckelberg type massive dark
photon $A'_\mu$ and the Standard Model photon $A_\mu$, given by
$\mathcal{L}\supset \varepsilon F'_{\mu\nu}F^{\mu\nu}/2$.
For small couplings, $\varepsilon\ll 1$, the
superradiance instability and subsequent quasi-monochromatic GW emission
proceeds as discussed above. On the other hand, for sufficiently large kinetic
mixing, a pair-production cascade ensues, filling the superradiance cloud with
a pair plasma, which continuously radiates electromagnetic
emissions with luminosities of up to $10^{43}$ erg/s \cite{Siemonsen2022}. 
Therefore, an electromagnetic counterpart is in principle expected.
However, typical binary mergers are likely outside the reach of multi-messenger searches
(see Ref.~\cite{Siemonsen2022} for further details).

Following Ref.~\cite{Siemonsen2022}, the cloud's dynamics can be classified
into three qualitatively different regimes depending on the mixing strength:
(i) $\varepsilon\ll \varepsilon_c \ll 1$, with critical kinetic mixing threshold 
$\varepsilon_c$, where the superradiance instability
and saturation through GW emission are essentially unaffected and the constraints obtained
in the standard scenario (as described in earlier sections and in
Ref.~\cite{Jones2023}) apply directly, (ii) $\varepsilon \sim \varepsilon_c$, where the cloud
evolution after gravitational saturation is impacted perturbatively, and (iii)
$\varepsilon\gg \varepsilon_c$, where the cloud's evolution after gravitational
saturation is driven by electromagnetic emission and hence impacted
non-perturbatively.\footnote{Notice, in all cases of relevance for CBC remnant
BH follow-up searches and for the regime of kinetic mixing that has not been
ruled out, the electromagnetic emission has no impact on the growth phase of
the superradiance cloud \cite{Siemonsen2022}.} In scenario (iii), the GW signal
sourced by the dark photon cloud is altered drastically compared to the purely
gravitational case, and the standard search methods are insensitive in those
regions of the dark photon parameter space. Therefore, we primarily focus on
determining the critical kinetic mixing $\varepsilon_c$, below which the
follow-up searches described in Sec.~\ref{sec:search_method} are sensitive to the
superradiant GW signal.

\begin{figure}
    \centering
    \includegraphics[width=1\linewidth]{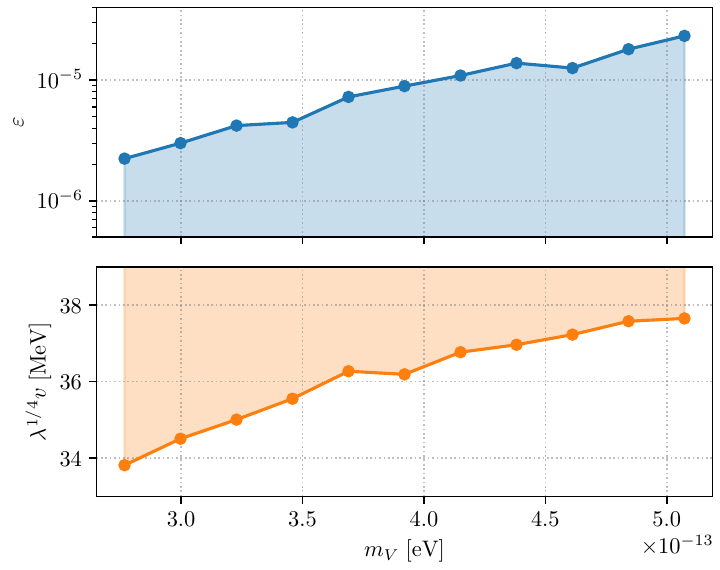}
    \caption{Parameter space of the kinetically mixed dark photon (top) and dark Higgs-Abelian sector (bottom) accessible by GW searches following up the GW170814-like event discussed in detail in Sec.~\ref{sec:synthetic_sys}. Points indicate the maximally conservative critical kinetic mixing $\varepsilon_c$ (top) and (fourth root of the) coupling $\rho_r$ (bottom); the shaded regions are those that could be excluded at the confidence level shown in Fig.~\ref{fig:Pdet_v_mV}.}
    \label{fig:darksynth}
\end{figure}

In case (ii), compared to the standard gravitational scenario, the GW emissions from the dark photon cloud are altered primarily in two ways: the signal amplitude decreases more quickly, and the frequency drift is larger. The strength of these effects is governed by the kinetic mixing, such that with $\varepsilon > \varepsilon_c$ the default search configuration can no longer track the signal's evolution. Thus, constraints obtained from null results in the standard scenario remain valid only for $\varepsilon<\varepsilon_c$.\footnote{Note that in general, $\varepsilon_c$ depends on $\alpha$ and the BH's spin.} In Appendix~\ref{app:cloudevo}, we describe in detail how $\varepsilon_c$ is obtained. In the top panel of Fig.~\ref{fig:darksynth}, we show the parameter space accessible with events similar to the synthetic GW170814-like event discussed in Sec.~\ref{sec:synthetic_sys}.
For each $m_V$, we determine the smallest $\varepsilon_c$ allowed by all posterior samples with a recovery rate of $0.9$ or above. Similar to the approach taken in Appendix~\ref{app:cloudevo}, this is the maximally conservative estimate. In Sec.~\ref{sec:popproj}, we discuss how the accessible parameter space shown in Fig.~\ref{fig:darksynth} compares to existing constraints.

\subsection{Dark Higgs-Abelian sector}
\label{sec:higgs}

We now turn to the scenario where the massive vector field is obtained from a
hidden U(1) gauge field $A'_\mu$ through a Higgs mechanism:
$\mathcal{L}=-F'_{\alpha\beta}F'^{\alpha\beta}/4-|D\Phi|^2/2-V(|\Phi|)/2$, where $D_\alpha=\nabla_\alpha-igA'_\alpha$ is the gauge covariant derivative
with gauge coupling $g$, and $V(|\Phi|)=(\lambda/2)(|\Phi|^2-v^2)^2$ is the
Higgs-like potential for the  complex scalar $\Phi$ with quartic coupling $\lambda$ and vacuum expectation value $v$. We
assume there are no further couplings of this dark sector to the Standard
Model. It has been shown that, depending on the relevant coupling strengths,
the presence of the Higgs boson may lead to an additional dark emission channel
\cite{Fukuda2019} or an explosive bosenova forming strings in the process
\cite{East2022,East2022_2} (see also Ref.~\cite{Brzeminski2024}). Both these
effects impact the cloud's evolution and GW emission and are controlled by the
critical amplitude $A'^2_c=\lambda v^2/g^2$, or equivalently the coupling $\rho=\lambda v^4$. Here, we also find that an overall shift in the superradiant cloud's energy levels can be 
induced by the presence of the Higgs.
In the parameter region of interest for this
work, i.e., $\alpha\sim 0.1$, we find the dark emission channel to be subdominant
during the nonlinear saturation process of the instability as shown explicitly in Appendix~\ref{app:cloudevo}.
Hence, the superradiant growth is halted by the spin-down of the BH or, for sufficiently
small couplings $\rho$, by strongly nonlinear string production within the cloud. The
shift in energy levels and dark bosonic radiation only impact the 
frequency evolution after gravitational saturation. Therefore, to determine the
applicability of null results obtained from a standard GW search following up
CBC remnant BHs (which assume a massive Stueckelberg vector), both string
production and the altered frequency evolution must be taken into account.

We first consider the impact of string production. String production commences spontaneously once the superradiant cloud's field
amplitude $A'^2$ approaches $A'^2_c$ \cite{East2022}. 
If strings are produced, the superradiance instability saturates at lower amplitudes, lowering
the emitted GW luminosity and likely altering the frequency evolution significantly 
(see also Ref.~\cite{Brzeminski2024});
this means the GW emission is no longer quasi-monochromatic. 
Therefore, search methods for such GWs from
the superradiant cloud in the standard scenario, as well as the derived
constraints, are applicable only if $A'^2$ remains below a critical amplitude,
which we denote as $A'^2_s$ (with $A'^2_s < A'^2_c$ as we demonstrate in
Appendix~\ref{app:cloudevo}),
throughout its evolution.\footnote{Strings may also be produced through other mechanisms
(e.g., the Kibble mechanism active in the early universe \cite{Kibble1980}) and
attach to the remnant BH \cite{Xing2020,East2022_2}. This could halt the
superradiance process and falsify the interpretation of a null
result from a GW search as the absence of such particles. We assume that the
remnant's string capture probability is negligible on timescales relevant for
follow-up searches.} Equivalently, this corresponds to couplings $\rho>\rho_s=m_V^2A'^2_s$. 
Systems satisfying this bound always
saturate gravitationally, and the subsequently emitted GWs are quasi-monochromatic.
At this stage, the presence of the Higgs boson impacts the GWs only perturbatively
through the dark bosonic radiation channel and shift in cloud energy levels; 
this is analogous to the kinetic mixing scenario considered above. 
Hence, if the coupling $\rho$ is above a threshold $\rho_r$, such that this impact is sufficiently small, then the standard GW search methods outlined above would still
be able to recover the signal just as in the pure gravitational case.
In Appendix~\ref{app:cloudevo}, we determine these critical
amplitudes and discuss associated theoretical uncertainties.
Notably, the coupling $\rho_r$, determined in Appendix~\ref{app:cloudevo}
and used below, is to be understood as an order-of-magnitude estimate.
Independently of these theoretical uncertainties, in the parameter space
relevant here, we find that the impact of the altered GW frequency evolution outweighs those
emerging from the string formation, i.e., $\rho_r>\rho_s$. 
All in all, null results of CBC remnant follow-up 
searches can be interpreted as constraints on a dark Higgs-Abelian sector
with couplings $\lambda v^4\gtrsim \rho_r$ for a given BH and boson mass.

Applying these conditions to the synthetic GW170814-like system considered above, we show in the bottom panel of Fig.~\ref{fig:darksynth} the dark Higgs-Abelian
parameter space accessible by current-generation ground-based GW detectors. 
This parameter space could be extended down to $\lambda^{1/4}v=\rho_s^{1/4}\lesssim\mathcal{O}(10)$ MeV
by decreasing the coherent time of the follow-up search configuration slightly.
Analogous to the case of the
kinetically mixed dark photon, we make the maximally conservative assumption
maximizing the critical field amplitudes (see Appendix~\ref{app:cloudevo} for 
further details) for a given $m_V$ over all posterior samples with recovery 
rate of $0.9$ and above.

\subsection{Next-generation detector projections}
\label{sec:popproj}

\begin{figure*}[t]
\includegraphics[width=0.5\textwidth]{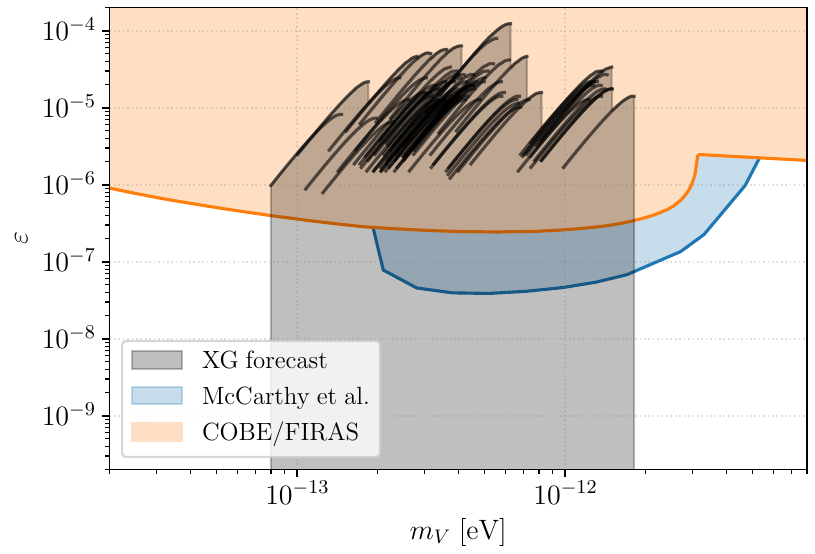}
\hfill
\includegraphics[width=0.49\textwidth]{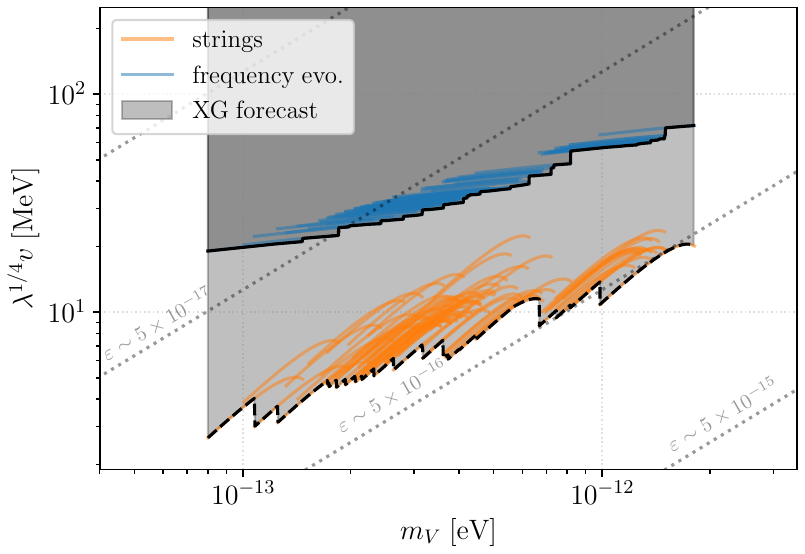}
\caption{Parameter space accessible by GW searches
(as outlined in Sec.~\ref{sec:method}) following up a population of detectable CBC
events using next-generation ground-based GW detectors
(labelled ``XG forecast''). Left: the kinetically mixed dark photon parameter space. Each black curve corresponds to a single such CBC event. 
Existing constraints, shown in blue and orange, were obtained in
Refs.~\cite{Mirizzi2009, Caputo2020} using data from COBE/FIRAS \cite{Fixsen1996}, as
well as by McCarthy et al. in Refs.~\cite{McCarthy2024, Pirvu2023} (see also
Ref.~\cite{Aramburo-Garcia2024}). Right: the dark Higgs-Abelian
model parameter space. In dark gray bounded by the solid black curve, we show the projected accessible parameter space. With slight modifications to the search algorithm, e.g., decreasing $T_{\rm coh}$ in the search configuration, the accessible parameter region could be extended down to the black dashed curve.
For each CBC event, we show the critical couplings $\rho_s^{1/4}$ (labelled ``strings'') 
and $\rho_r^{1/4}$ (labelled ``frequency evo.'') as one orange and one blue curve, respectively. The dashed and solid
curves combine the former and latter across events, respectively. 
The vector's mass is $m_V=gv$. The shapes of each individual curve (both on the left and right) is explained in detail in Appendix~\ref{app:cloudevo}. On the right, we also show the approximate kinetic mixing coefficient that would correspond to the Higgs-Abelian parameters under the assumption that it arises from loops of heavy fermions with order unity charge $\varepsilon\sim e g/(16 \pi^2)$~\cite{Holdom1986}. The dotted lines denote constant $\varepsilon/\lambda^{1/4}$ (or equivalently, constant $\varepsilon$ assuming $\lambda=1$).}
\label{fig:pop_ex}
\end{figure*}

Finally, we forecast the parameter spaces of a kinetically mixed dark
photon and a dark Higgs-Abelian model that are accessible by the CBC follow-up
search discussed above using next-generation ground-based GW
detectors, e.g., the Einstein Telescope and Cosmic Explorer. To that
end, we consider the currently detected population of $\lesssim
\mathcal{O}(100)$ (see Refs.~\cite{GWTC-1,GWTC-2,GWTC-3}) CBC
remnants.\footnote{We implicitly assume that these samples are a good
representation of the underlying CBC remnant population detectable by
next-generation detectors.} We make use of the horizon distance $d_H$ derived
in Ref.~\cite{Jones2023}, i.e., the luminosity distance out to which a signal
from a given remnant BH, in conjunction with its optimally matched boson
$m_V^{\rm opt}$, can be detected given next-generation detector sensitivities 
(at $95\%$ confidence for a $1\%$ false alarm probability).
For each set of CBC remnant properties ($M$, $\chi$, and luminosity distance
$d$), we determine whether the remnant BH is within reach, i.e., $d<d_H$. The
boson mass range, which can be probed with each given remnant, is determined
assuming that $\alpha\in [0.6,1.1]\alpha_{\rm opt}$ (see
Sec.~\ref{sec:simulations}). Finally, for any given remnant BH-boson system,
the critical kinetic mixing $\varepsilon_c$, as well as the critical 
couplings $\rho_r$ and $\rho_s$, are determined using the results from Appendix~\ref{app:cloudevo}.

In Fig.~\ref{fig:pop_ex}, we show the parameter spaces of a kinetically mixed
dark photon and a dark Higgs-Abelian model that can be probed by
next-generation ground-based GW detectors. In the accessible dark photon mass
range, the \textit{entire remaining} parameter space, unconstrained by cosmic 
microwave background observations, can be probed by
following up a few CBC remnants. Additionally, these constraints in the 
ultralight mass range can be mapped to bounds down to $\lambda^{1/4}v \sim \mathcal{O}(10)$ MeV 
on a dark Higgs-Abelian sector. Importantly, in the latter case, existing 
constraints on ultralight vector bosons obtained from spin measurements and 
GW searches likely already apply down to a similar level of $\lambda^{1/4}v$ 
(though with the attendant assumptions commented on in Sec.~\ref{sec:introduction}).
Notice, the accessible
boson mass ranges shown in Fig.~\ref{fig:pop_ex} are a conservative estimate:
the currently detected sample of CBC remnant BHs may not have explored the
tails of the underlying population. However, next-generation detectors will be sensitive
to heavier binary systems, and GW search techniques may be improved in the
future, so in fact the accessible mass range likely extends further towards higher
and lower dark photon masses. Lastly, for the BH remnant samples considered
here, both the Einstein Telescope and Cosmic Explorer are similarly sensitive.


\section{Conclusion}
\label{sec:conclusion}

In this paper, we build upon the vector boson search method laid out in Ref.~\cite{Jones2023}, describing a statistically robust procedure to carry out follow-up searches targeting CBC remnant BHs. We formulate the method to derive constraints on the vector boson mass in the absence of a detection by taking into account the uncertainties of the BH parameters. 
We demonstrate the procedure using a synthetic GW170814-like system as an example. We also describe how to interpret vector mass constraints when considering an extended dark sector such as the dark photon or the dark Higgs-Abelian sector. In particular, we show the constraints on the vector mass derived in the standard scenario remain valid in these models up to a certain coupling strength between the vector field and the dark sector.

Although we used real detector noise ASD estimates when running simulations with the synthetic system (Sec.~\ref{sec:synthetic_sys}), for simplicity, we did not factor in other data quality issues that might arise in real searches such as data gaps, glitching artifacts, etc. Our simulations can therefore be viewed as the ideal case. 
Fortunately, the search pipeline described in this paper is well-equipped to handle data gaps. The general procedure is to assign equal likelihoods in all the frequency bins during the time in which there is no available observational data. 
However, the existence of data gaps reduces the search sensitivity due to the reduced SNR one could accumulate. 

For the synthetic system studied in Sec.~\ref{sec:synthetic_sys}, we could exclude at best a fraction of one order of magnitude in the full vector boson mass range, assuming it was a real event and the search did not yield a detection.
In reality, most CBC remnant BHs to date are still beyond reach for vector searches (or marginally reachable but not promising enough to derive constraints with high confidence).
In future observing runs with further improved detectors, and in the era of next-generation detectors, we can expect a large population of CBC detections with much higher SNRs~\cite{GWTC-3, GW_observation_prospects}, which will enable sensitive searches for vector boson clouds hosted by the remnants.
We also mention that recently an improved model for the frequency evolution of
the GW signals from vector clouds has been obtained in
Ref.~\cite{May2024}. Though the slightly increased estimates for the frequency
time derivative obtained in this improved model (by $\sim 10\%$ for the typical systems
considered here) will not affect the conclusions of this work, a more accurate frequency evolution model could not only be used to more optimally tune
the configuration of the search strategy used here, but might also be leveraged
to design more sensitive searches that track the signal coherently over longer
timescales.
By targeting remnant BHs with a variety of masses, one can correspondingly probe a wide boson mass range.
Thus, this work merely lays the groundwork for the broader constraints that could be derived with future GW observations in a higher SNR regime.

We derive explicit expressions that map constraints on a vector mass range also to constraints on a
kinetically mixed dark photon as well as a dark Higgs-Abelian sector. Applying these mappings to the synthetic GW170814-like remnant BH and a larger population of CBC remnants, we find that if confident constraints on ultralight vector bosons can be placed, the dark photon parameter space unconstrained by cosmic microwave background observations up to kinetic mixings of $\varepsilon\sim 10^{-6}$ can be accessed.
Analogously, ultralight vector constraints apply down to Higgs couplings of $\lambda^{1/4} v\sim\mathcal{O}(10)$ MeV. While we apply the mappings explicitly to possible constraints obtained from GW searches following up CBC remnants, these mappings may also apply to other existing constraints obtained from searches for signatures of superradiant clouds (in particular in the case of a dark Higgs-Abelian sector), such as spin measurements and other GW searches.

\begin{acknowledgments}
We thank Junwu Huang and Masha Baryakhtar for many insightful discussions.
The authors are grateful for computational resources provided by the LIGO Laboratory and supported by National Science Foundation Grants PHY-0757058 and PHY-0823459.
This research is supported by the Australian Research Council Centre of Excellence for Gravitational Wave Discovery (OzGrav), Project Numbers CE170100004 and CE230100016. L.S. is also supported by the Australian Research Council Discovery Early Career Researcher Award, Project Number DE240100206.
W.E. acknowledges support from a Natural Sciences and Engineering Research
Council of Canada Discovery Grant and an Ontario Ministry of Colleges and
Universities Early Researcher Award. This research was supported in part by
Perimeter Institute for Theoretical Physics. Research at Perimeter Institute is
supported in part by the Government of Canada through the Department of
Innovation, Science and Economic Development and by the Province of Ontario
through the Ministry of Colleges and Universities. 
This research was enabled in part by support
provided by SciNet (www.scinethpc.ca) and the
Digital Research Alliance of Canada (alliancecan.ca). 
This research has made use of data or software obtained from the Gravitational Wave Open Science Center (gwosc.org), a service of the LIGO Scientific Collaboration, the Virgo Collaboration, and KAGRA. This material is based upon work supported by NSF's LIGO Laboratory which is a major facility fully funded by the National Science Foundation, as well as the Science and Technology Facilities Council (STFC) of the United Kingdom, the Max-Planck-Society (MPS), and the State of Niedersachsen/Germany for support of the construction of Advanced LIGO and construction and operation of the GEO600 detector. Additional support for Advanced LIGO was provided by the Australian Research Council. Virgo is funded, through the European Gravitational Observatory (EGO), by the French Centre National de Recherche Scientifique (CNRS), the Italian Istituto Nazionale di Fisica Nucleare (INFN) and the Dutch Nikhef, with contributions by institutions from Belgium, Germany, Greece, Hungary, Ireland, Japan, Monaco, Poland, Portugal, Spain. KAGRA is supported by Ministry of Education, Culture, Sports, Science and Technology (MEXT), Japan Society for the Promotion of Science (JSPS) in Japan; National Research Foundation (NRF) and Ministry of Science and ICT (MSIT) in Korea; Academia Sinica (AS) and National Science and Technology Council (NSTC) in Taiwan.
\end{acknowledgments}  

\appendix

\section{Search configurations}
\label{appendix:configs}

\begin{figure}[hbt!]
    \centering
    \includegraphics[scale=.42]{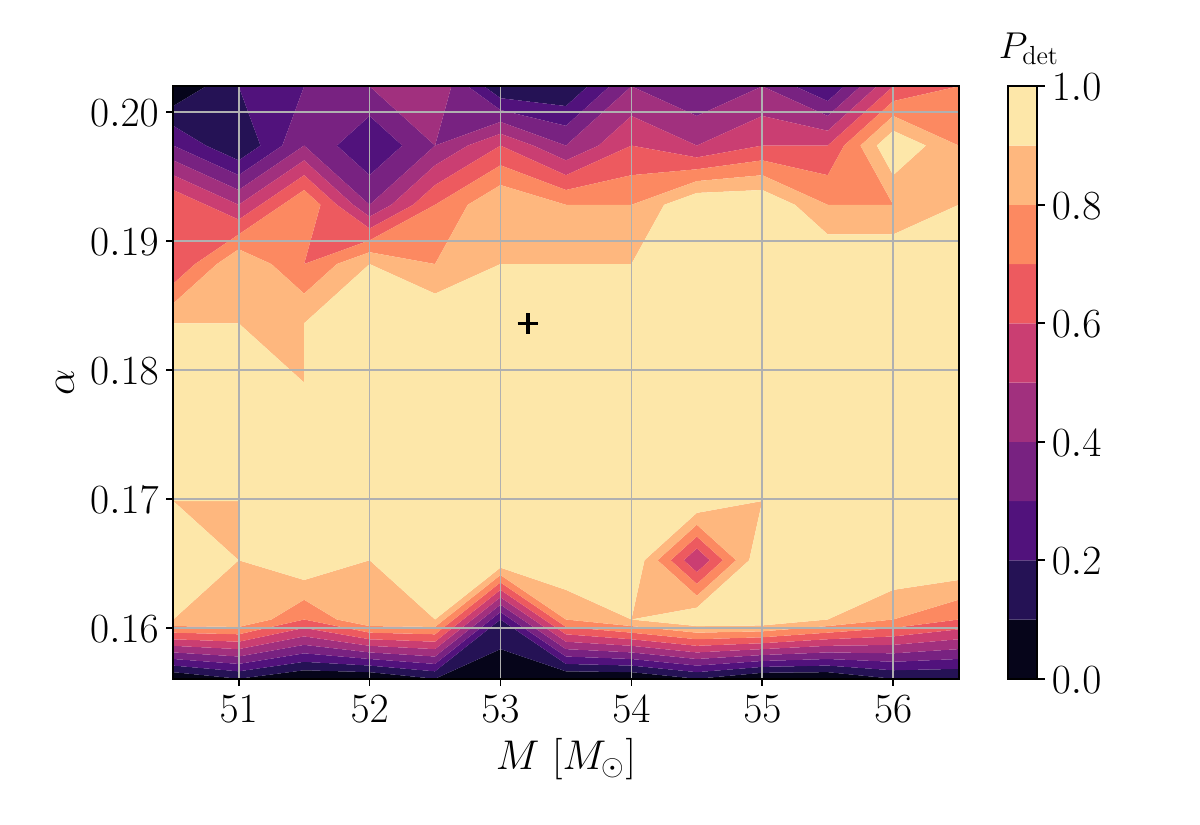}
    \caption{Detection probability $P_{\rm det}$ for synthetic systems with a grid of parameters of $M$ and $\alpha$ (with all other BH parameters fixed to the median values shown in Table~\ref{tab:GW170814}). The search configuration is fixed at $t_{\rm start} - t_0 = 3.62$~hr, $T_{\rm coh} = 7.2$~min, and $T_{\rm obs} = 9.72$~hr. The plus marker indicates the BH-boson system with $M = 53.2~M_{\odot}$ and $\alpha = 0.184$, i.e., the system for which the selected configuration is optimal.
}
    \label{fig:recovery_rate_fixed_config}
\end{figure}

For the search and follow-up study described in Secs.~\ref{sec:method}--\ref{sec:simulations}, one must choose a set of configurations that is both computationally feasible and provides sufficient coverage of the full parameter space. Because we use a flexible HMM-based search method and only track the signal frequency evolution over time, we find that a single configuration (with parameters $t_{\rm start}$, $T_{\rm coh}$, and $T_{\rm obs}$) is able to recover signals from systems with a range of parameters.
Note, in this work we consider a signal detectable for a given configuration if, after running multiple noise realizations, the recovery rate is $\geq 90\%$, i.e., the false dismissal probability is $\leq 10\%$. An example is given in Fig.~\ref{fig:recovery_rate_fixed_config}, where the detection probability is shown as the colored contours for synthetic injections generated by systems with a grid of parameters $M$ and $\alpha$ (all other parameters fixed to the medians shown in Table~\ref{tab:GW170814}). The search configuration, optimal for a system with parameters marked by ``$+$'' in the figure, remains fixed across all injections: $t_{\rm start} - t_0 = 3.62$~hr, where $t_0$ is the time at which the BH was born, $T_{\rm coh} = 7.2$~min, and $T_{\rm obs} = 9.72$~hr. This single configuration is able to recover injected signals across the majority of the parameter space (with a 1\% false alarm probability), even if, in most cases, it is not the optimal configuration.\footnote{The optimal configuration refers to $t_{\rm start} = t_{\rm sat}$ (the saturation time of the superradiant cloud growth), $T_{\rm coh} \approx (2 \dot{f}^{\rm max}_{\rm GW})^{-1/2}$ [see Eq.~\eqref{eqn:Tcoh}], and $T_{\rm obs} \sim \tau_{\rm GW}$ (the GW emission timescale)~\cite{Jones2023}.} 
As demonstrated in Sec.~\ref{sec:simulations}, such a synthetic system is placed just within reach for us to derive meaningful constraints on $m_V$. The SNRs of the simulated signals are not so high that any mismatched search configuration would have recovered them. Hence, in practice, for any target system that our search is sensitive to, it is generally the case that a single search configuration can cover at least a range of the relevant parameter space. 

Nevertheless, the parameter space considered in a real vector boson search is usually larger than that shown in the above example. Given the uncertainties associated with the remnant BH properties and the unknown boson mass range (or equivalently, the range of $\alpha$), it is beneficial to use multiple search configurations to provide better coverage and improve the overall search sensitivity. Still, one must be conservative in selecting the configurations.
Here we describe the procedure of selecting a set of search configurations for a given target, using the synthetic GW170814-like system discussed in Sec.~\ref{sec:simulations} as an example. It is worth noting, however, that these guidelines should be generally applicable to the majority of merger remnants one might target in a vector boson search---that is, there is nothing special about the GW170814-like system we consider here.

\begin{figure}[hbt!]
    \centering
    \includegraphics[scale=.5]{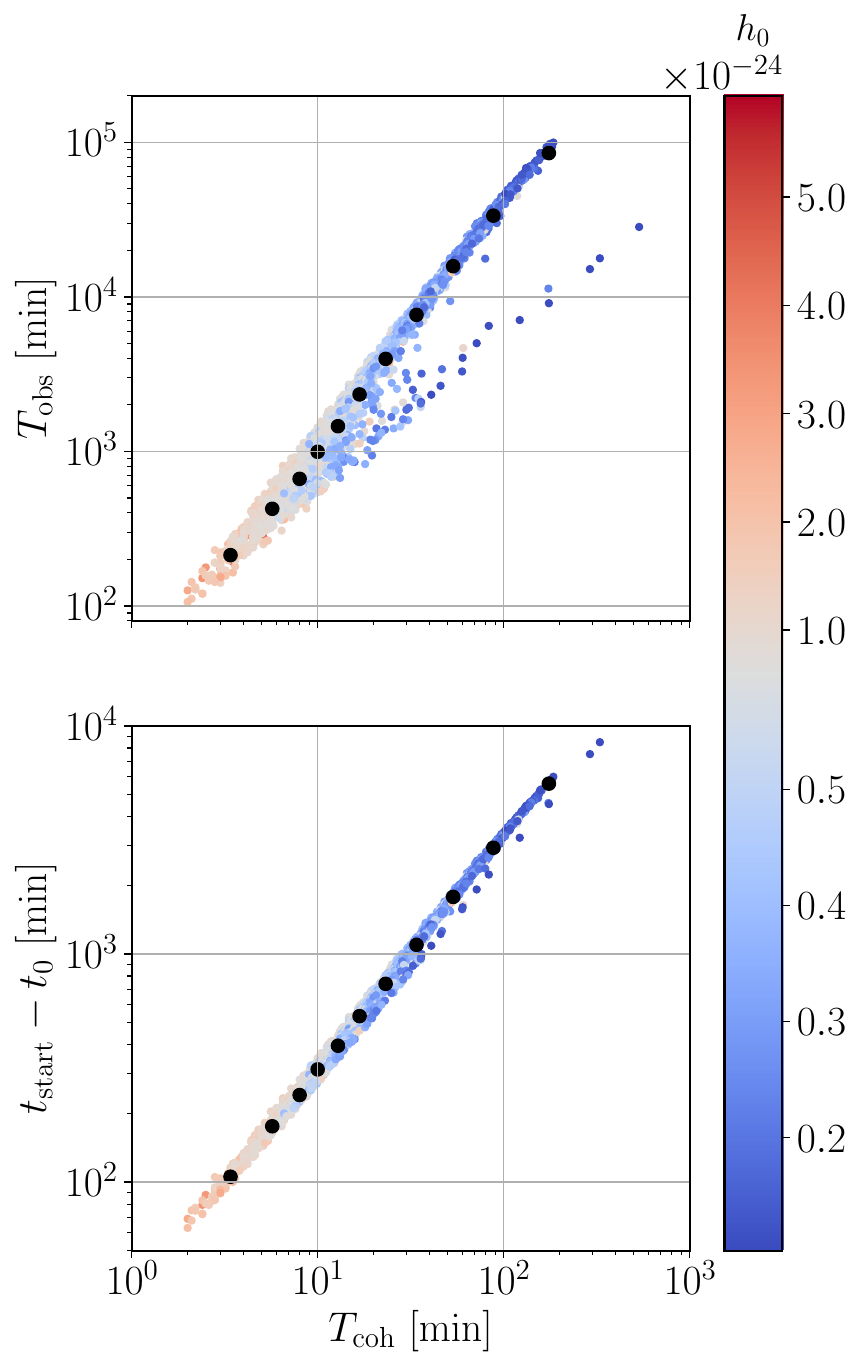}
    \caption{Optimal search configuration parameters \{$t_{\rm start} - t_0, T_{\rm coh}, T_{\rm obs}$\} for each of $N_{\rm BH}=200$ sample BHs with each of the $11$ $m_{V}$ values. The color corresponds to the predicted characteristic strain amplitude $h_0$ of each system at $t_{\rm sat}$. The black dots represent 11 configurations chosen to cover the full parameter space. The top and bottom panels show $T_{\rm obs}$ and $t_{\rm start} - t_0$ versus $T_{\rm coh}$, respectively.}
    \label{fig:search_parameters}
\end{figure}

We first calculate the optimal search configuration for each of the $N_{\rm BH} = 200$ sample BHs and each of the 11 $m_V$ values we consider. 
The optimal configuration ($t_{\rm start}$, $T_{\rm coh}$, and $T_{\rm obs}$) for each single system is plotted as a colored dot in Fig.~\ref{fig:search_parameters}.
The color represents the predicted characteristic strain amplitude $h_0$ of each system at $t=t_{\rm sat}$. 
In each panel, two out of the three configuration parameters are shown. Since all three parameters depend on the cloud growth and emission timescales, there is an approximately linear relationship between any two parameters. Clouds growing more slowly emit longer signals, and thus configurations with later $t_{\rm start}$ and longer $T_{\rm coh}$ and $T_{\rm obs}$ can achieve better sensitivities. 
To ensure the search configurations adequately cover the full range of system parameter uncertainties, we take values at 11 percentiles in the $t_{\rm start}$, $T_{\rm coh}$, and $T_{\rm obs}$ distributions (drawn independently, not jointly) and form 11 sets of search configurations, listed in Table~\ref{tab:configs}. The 11 search configurations are shown as black dots in Fig.~\ref{fig:search_parameters}. The black dots in the top and bottom panels are from the same 11 configurations but with different parameters displayed.
The percentile spacing is chosen conservatively such that most sample systems can be recovered by more than one configuration. At the same time, we find that using more than 11 configurations does not further improve the detection probabilities quoted in Fig.~\ref{fig:Pdet_v_mV} and thus is not worth the additional computational expense.

\begin{table}[tbh!]
    \centering
    \setlength{\tabcolsep}{8pt}
    \renewcommand\arraystretch{1.2}
    \begin{tabular}{lccc}
        \hline
    Percentile & $t_{\rm start} - t_0$ [min] & $T_{\rm coh}$ [min] & $T_{\rm obs}$ [min] \\
        \hline
    2 & 105.6 & 3.4 & 210.8 \\
    10 & 175.8 & 5.6 & 420.0 \\
    20 & 240.9 & 8.0 & 664.0 \\
    30 & 312.1 & 10.0 & 990.0 \\
    40 & 396.2 & 12.8 & 1446.4 \\
    50 & 534.2 & 16.8 & 2335.2 \\
    60 & 739.9 & 23.2 & 3967.2 \\
    70 & 1097.3 & 34.0 & 7616.0 \\
    80 & 1778.9 & 53.4 & 15753.0 \\
    90 & 2921.4 & 88.0 & 33528.0 \\
    98 & 5581.1 & 175.0 & 85225.0 \\
        \hline
    \end{tabular}
    \caption{Search configuration parameters and the percentiles at which they are drawn (shown as black dots in Fig.~\ref{fig:search_parameters}).}
    \label{tab:configs}
\end{table}

\section{Details on mapping constraints to larger dark sector}
\label{app:cloudevo}

In the following, we outline how a constraint on the existence of a vector boson in a specific mass range can be mapped directly to constraints on kinetically mixed dark photons and a dark Higgs-Abelian sector. The search methods developed above and in Ref.~\cite{Jones2023} are tuned towards the emission of long-transient GWs from vector boson clouds, ignoring any non-gravitational interaction. Despite this, the search algorithm is sufficiently flexible to allow for mild modifications of the gravitational waveform due to non-vanishing non-gravitational interactions. In this section, we quantify precisely how large these modifications may be and beyond what level the signal tracking becomes unreliable. More concretely, we determine the critical kinetic mixing $\varepsilon_c$ and critical amplitudes $A'^2_r$ and $A'^2_s$, defined in the main text, to which any constraint on the vector's mass derived in the pure gravitational scenario can be applied (at the same confidence). For notational ease, we set $G=1$ for the remainder of the appendix. This does not impact any quantity quoted from this appendix in the main text.

\subsection{Kinetically mixed dark photon}

\subsubsection{Modified cloud evolution}

Recalling the results of Ref.~\cite{Siemonsen2022}, in the presence of a non-vanishing kinetic mixing, the superradiant cloud dissipates energy through the emission of Standard Model electromagnetic radiation. The relevant timescales are the GW emission timescale $\tau_{\rm GW}=(M_c/\dot{E}_{\rm GW})|_{t=t_{\rm sat}}$ (defined using the GW energy flux $\dot{E}_{\rm GW}$ and the cloud's mass $M_c$ at the instability saturation time), the electromagnetic emission timescale $\tau_{\rm EM}=M_c/L_{\rm EM}$ (which depends on the electromagnetic energy flux $L_{\rm EM}$), and the superradiance energy extraction timescale $\tau_{\rm SR}$. The energy flux through the horizon during the evolution of the superradiance instability (for the fastest growing field configuration) scales as $\dot{E}_{\rm BH}=\tau_{\rm SR}^{-1} M_c$ with $\tau_{\rm SR}^{-1}=4\alpha^7\chi M$ in the $\alpha\ll 1$ limit \cite{Baryakhtar2017}. Therefore, for any $\varepsilon$, there exists a sufficiently small $\alpha$ such that $\tau_{\rm SR}>\tau_{\rm EM}$. In this regime, the cloud saturates as soon as the pair production rate is efficient \cite{Siemonsen2022}, which typically happens \textit{before} the gravitational saturation, when $\omega\approx\Omega_H$. However, for reasonable kinetic mixing parameters, this occurs deep in the non-relativistic regime; for instance, for $\varepsilon=10^{-6}$, the critical fine structure constant below which the above situation becomes relevant is $\alpha_c\approx 5 \times 10^{-3}$ \cite{Siemonsen2022}. Hence, in the following, we assume the superradiance instability saturates through either gravitational or electromagnetic emissions at $\omega=\Omega_H$, i.e., $\tau_{\rm EM}\gg\tau_{\rm SR}$.

In this case, the cloud's mass evolution after saturation follows 
\begin{align}
\dot{M}_c=-L_{\rm EM}-\dot{E}_{\rm GW}.
\label{eq:Mcevolution}
\end{align}
The electromagnetic energy flux of the fastest growing field configuration is given by \cite{Siemonsen2022}
\begin{align}
L_{\rm EM}= & \ \varepsilon^2 F(\alpha)\frac{M_c}{M}, \\
F(\alpha)= & \ 0.131\alpha-0.188\alpha^2
\label{eq:EMluminosity}
\end{align}
and is valid up to $\alpha\lesssim 0.5$. For small $\alpha$, the GW energy flux scales as $\dot{E}_{\rm GW}\sim \alpha^{10} M_c^2/M^2$ \cite{Baryakhtar2017,Siemonsen2020}, whereas for $\alpha\gtrsim 0.1$ (i.e., the parameter regime relevant for this work) the flux can be determined only numerically \cite{Siemonsen2020}; we make use of \texttt{SuperRad} to compute $\dot{E}_{\rm GW}$ in this regime. The solution to \eqref{eq:Mcevolution} is, in general,
\begin{align}
M_c(t)=\frac{M_c^{\rm sat}\tau_{\rm GW}}{\tau_{\rm GW}e^{t/\tau_{\rm EM}}+\tau_{\rm EM}(e^{t/\tau_{\rm EM}}-1)},
\label{eq:Mcexplicit}
\end{align}
where we chose $t=0$ to correspond to the saturation with $M_c(0)=M_c^{\rm sat}$. Controlled by the ratio of the energy fluxes, $\beta\equiv\tau_{\rm GW}/\tau_{\rm EM}$, there are three qualitatively different regimes. In the two extreme limits, the cloud's evolution is either dictated entirely by GW emission,
\begin{align}
    M_c(t)=\frac{M_c^{\rm sat}}{1+t/\tau_{\rm GW}}, & & \text{for} \ \beta\ll 1
\end{align}
or entirely by electromagnetic radiation,
\begin{align}
    M_c(t) = M_c^{\rm sat}e^{-t/\tau_{\rm EM}}, & & \text{for} \ \beta\gg 1.
\end{align}
For $\beta\sim\mathcal{O}(1)$, both dissipation channels are relevant, and the cloud's evolution is given by \eqref{eq:Mcexplicit}. The parameter $\beta$ is implicitly a function of $\varepsilon$, $\alpha$, and the BH's spin. 

\subsubsection{Critical kinetic mixing}

For a CW search, the most important quantities are the GW amplitude and frequency shift, which are directly proportional to the cloud's mass and its time derivative: $h_0\propto M_c(t)$ and $\dot{f}_{\rm GW}(t)\propto \dot{M}_c(t)$. More concretely and following Ref.~\cite{Jones2023}, given a remnant BH mass and spin, as well as a boson mass, the search techniques introduced in Sec.~\ref{sec:search_method} are primarily configured by a choice of $T_{\rm coh}$ and $T_{\rm obs}$; these are chosen directly based on the GW emission timescale $\tau_{\rm GW}$ and frequency derivative at saturation $\dot{f}_{\rm GW}(t=0)$, assuming $\varepsilon=0$. Therefore, null results of these searches can be interpreted as constraints on a kinetically mixed dark photon, if the mixing is below a critical value $\varepsilon_c$, such that the choices of $T_{\rm coh}$ and $T_{\rm obs}$ suffice for the signal to be successfully tracked and lie above the detection threshold. This leads us to impose two conditions for the applicability of constraints on the dark photon parameter space.

First, an injection of a GW signal from a cloud around the fiducial BH, with remnant mass $M=60\ M_\odot$, dimensionless spin $\chi=0.7$, fine structure constant $\alpha=0.176$, and luminosity distance $d=500$ Mpc, can be marginally recovered by search methods introduced in Sec.~\ref{sec:search_method} for $T_{\rm obs}\approx 2\tau_{\rm GW}$ \cite{Jones2023}. Therefore, the signal is still marginally detectable even if the cloud mass decays faster after saturation due to the additional emission channel, with the initial half-life of the cloud being half of the assumed $\tau_{\rm GW}$, i.e., $M_c(t=\tau_{\rm GW}/2)=M_c^{\rm sat}/2$. Given the explicit form \eqref{eq:Mcexplicit}, this condition is equivalent to finding the solution $\beta_1\approx 0.68$ to the transcendental equation
\begin{align}
\frac{2\beta_1}{e^{\beta_1/2}(1+\beta_1)-1}=1.
\end{align}
That is, all signals with $\beta \lesssim \beta_1$ are detectable if $T_{\rm obs}$ is chosen based on the $\varepsilon=0$ expectations for the above fiducial BH remnant. In principle, a larger injection study would be required to confirm this more broadly. However, as shown below, this is the less conservative condition on the kinetic mixing strength, so we refrain from investigating this condition across a larger set of search configurations.

Second, for successful tracking of a signal assuming vanishing kinetic mixing, the frequency must not wander out of the frequency bin $\Delta f=1/(2T_{\rm coh})$ corresponding to the chosen coherent time $T_{\rm coh}$:
\begin{align}
    \Delta f\geq f_{\rm wander}^{\beta=0}(t)=\left|\int_t^{t+T_{\rm coh}}dt'\dot{f}^{\beta=0}_{\rm GW}(t')\right|.
\end{align}
For a given remnant BH and boson mass, the coherent time is chosen by neglecting any frequency derivative evolution and assuming the maximal possible frequency drift $\dot{f}^{\beta=0}_{\rm GW}(0)$, which is still marginally bounded by $\Delta f$; that is, $f_{\rm wander}^{\beta=0}(0)$ is approximated by $\dot{f}_{\rm GW}^{\beta=0}(0) T_{\rm coh}$, such that $T_{\rm coh}=[2 \dot{f}_{\rm GW}^{\beta=0}(0)]^{-1/2}$ \cite{Jones2023}. However, since $\dot{f}_{\rm GW}(t)$ monotonically decreases in time and $f_{\rm wander}(0) >f_{\rm wander}(t>0)$ (both for $\varepsilon=0$ and $\varepsilon\neq 0$), choosing $T_{\rm coh}$ based on the maximal possible frequency drift at $t=0$ means that the fixed $T_{\rm coh}$ over the total signal tracking duration is conservative for any time at $t>0$. Therefore, we define the critical flux ratio $\beta_2$ implicitly by considering the explicit time-dependence of the frequency derivative. To that end, we estimate the integral $f^{\beta=\beta_2}_{\rm{wander}}(0)$ using a trapezoidal approximation, 
\begin{align}
\begin{aligned}
    f^{\beta=\beta_2}_{\rm wander}(0)\approx & \  \dot{f}^{\beta=\beta_2}_{\rm GW}(0) T_{\rm coh} \\
    & \ -\frac{T_{\rm coh}}{2}\left[\dot{f}^{\beta=\beta_2}_{\rm GW}(0)-\dot{f}^{\beta=\beta_2}_{\rm GW}(T_{\rm coh}) \right]\\
    \equiv & \ f^{\beta=\beta_2}_{\rm{approx}},
\end{aligned}
\label{eq:frequencycondition}
\end{align}
and impose this approximation to agree with $\Delta f$ defined above: $f^{\beta=\beta_2}_{\rm{approx}}=\dot{f}_{\rm GW}^{\beta=0}(0) T_{\rm coh}=\Delta f$.
This condition reduces to $\beta_2\approx T_{\rm coh}/\tau_{\rm GW}\approx T_{\rm coh}/T_{\rm obs}=N_T^{-1}$ in the $T_{\rm coh}/\tau_{\rm GW}\ll 1$ limit. For all search configurations used here, this remains in the range $N_T\in [25, 500]$; we conservatively use $\beta_2= 10^{-3}$. Evidently, this is more conservative than $\beta_1$ independent of the BH remnant and boson mass. Therefore, all projections in the main text are based on $\beta_2$. Notice, in an actual follow-up search $N_T$ is known and the conditions on $\beta$ can be tightened. 

These conditions on the flux ratio of emitted gravitational and electromagnetic waves bound the dark photon parameter space, in which null results can be interpreted as constraints, to 
\begin{align}
\varepsilon \leq\varepsilon_c =\left(\frac{M}{\tau_{\rm GW}}\frac{\beta_2}{F(\alpha)}\right)^{1/2},
\end{align}
which follows directly from the definition of the flux ratio $\beta$. Note, $\tau_{\rm GW}/M$ is implicitly dependent on both $\alpha$ and $\chi$ and is estimated accurately by \texttt{SuperRad} across the parameter space.

\subsection{Dark Higgs-Abelian sector}

\subsubsection{Modified frequency evolution}

During the superradiance instability phase of the evolution, the cloud's density grows exponentially towards large field amplitudes. While at low densities, the underlying vector field behaves like a massive (i.e., Stueckelberg) vector field, the presence of the Higgs boson perturbatively manifests itself as effective higher-order self-interactions of this vector \cite{Fukuda2019}. We focus on those self-interactions arising from integrating out the radial mode of the Higgs; the angular mode likely leads to derivative self-interactions at low densities, which are high-order in $\alpha$. These self-interactions, quartic in nature at lowest order, can alter the bound state's frequencies and up-scatter these states into asymptotically free radiation (in direct analogy to the case of scalar self-interactions \cite{Arvanitaki2011,Baryakhtar2020,Omiya2022}). We quantify the self-interaction-induced change of the cloud's energy levels below. In Ref.~\cite{Fukuda2019}, the basic scaling of the energy flux associated with the asymptotically free radiation is determined to be $\dot{E}_{\rm rad}\sim \alpha^6 A'^{-4}_c(M_c/M)^3$ in the $\alpha\ll 1$ limit, where recall $A'^2_c=\lambda v^2/g^2$. However, the numerical coefficient preceding this scaling is unknown. Results presented in Ref.~\cite{East2022} can be used for a rough estimate; however, since the focus there lies on the strongly nonlinear string formation regime, the resulting fluxes are likely enhanced compared to the weakly nonlinear radiation emitted by quartic self-interactions. 

To obtain a more accurate estimate, albeit still with large uncertainties as
discussed below, we perform numerical time-domain simulations of the Higgs-Abelian
field equations on a Kerr BH background from appropriately chosen initial
data as in Ref.~\cite{East2022}. Here we focus on scenarios deep in the weakly 
nonlinear regime, i.e., at low
field amplitudes. First, these simulations confirm that $\dot{E}_{\rm rad}$ is
enhanced by higher-order nonlinearities as $A'^2$ approaches $A'^2_c$ before
strings are formed. Second, isolating the leading quartic contribution to
$\dot{E}_{\rm rad}$ (i.e., those that scale as $\sim M_c^3$), one finds a steeper
than $\sim \alpha^6$ scaling of the flux in the $\alpha\in(0.3,0.5)$ regime.
This suggests that extrapolating these results to $\alpha\approx 0.1$ (relevant for
this work) and using a $\sim \alpha^6$ scaling overestimates the flux in that
regime. Lastly, using our evolution at $\alpha=0.3$, we can place the upper
limit on the flux ratio, $\dot{E}_{\rm rad} A'^4_c/\dot{E}_{\rm BH}^3\lesssim 6$, 
where $\dot{E}_{\rm BH}$ is the bosonic flux through the BH horizon
(defined to be positive if energy is extracted from the BH, and which can 
be independently computed using \texttt{SuperRad}). Assuming
this upper limit can be extrapolated as $\propto\alpha^6$,
we obtain $\dot{E}_{\rm rad}\approx 3\times
10^{-9}\alpha^6 A'^{-4}_c(M_c/M)^3$. This suggests that $\dot{E}_{\rm rad}/\dot{E}_{\rm BH}<1$
for $A'^2\approx A'^2_c$ and $\alpha\sim 0.1$; hence, the superradiance instability saturates gravitationally.
Due to the uncertainties in our numerical time-domain
evolution, the large extrapolation from the relativistic to the
non-relativistic regimes, and the suggested steeper scaling with $\alpha$, this
estimate of the flux is to be understood as an (at best) order-of-magnitude estimate. For
the reasons above, this is likely a conservative estimate for $\alpha\lesssim
0.3$, i.e., it overestimates the total flux in the regime of parameter space
relevant to this work and, hence, the impact on the GW frequency evolution.

The impact on the bound state energy levels due to the effective self-interactions
can be estimated in the non-relativistic limit, i.e., for $\alpha\ll 1$, analogously to 
the shift due to the cloud's self-gravity \cite{Baryakhtar2020, Siemonsen2022_2}. 
Integrating out the
radial mode as done in Ref.~\cite{Fukuda2019}, the superradiant bound states
are subject to the perturbing potential $\delta V=-\mu A'^2/(2A'^2_c)$.\footnote{We use the mostly plus metric signature.} At leading
order in $\alpha$, the most unstable unperturbed bound state exhibits the
time-independent field amplitude $A'^2=\alpha^3\mu M_c \exp(-2\mu\alpha r)/\pi$. 
Using a calculation analogous to that presented in Ref.~\cite{Baryakhtar2020}, the shift in angular
frequency due to this self-interaction on the most unstable state is
\begin{align}
    \Delta\omega_{A'}=-\frac{\alpha^4\mu}{16\pi}\frac{M_c}{M} A'^{-2}_c.
    \label{eq:deltaomgeaA}
\end{align}
The corresponding self-gravity correction to the angular frequency (in the non-relativistic limit) is 
$\Delta\omega_G=-5\alpha^2\mu M_c/(8M)$ \cite{Siemonsen2022_2}. 
For notational convenience, we define the $\gamma\equiv g^2/(\lambda v^2)$. 
Hence, $\gamma=0$ corresponds to the purely gravitational (i.e., Stueckelberg) case, 
while $\gamma\neq 0$ implies a non-vanishing additional shift of the energy levels 
due to the dark Higgs.
The GW frequency derivative due to these two effects is then 
$\dot{f}^\gamma_{\rm GW}=(\Delta\omega_G+\Delta\omega_{A'})\dot{M}_c/(\pi M_c)$.

Given the additional radiation channel, the cloud's mass evolution after gravitational saturation follows
\begin{align}
    \dot{M}_c=-\dot{E}_{\rm GW}-\dot{E}_{\rm rad},
    \label{eq:bosonicrad}
\end{align}
to leading order in the self-interactions. Notably, in contrast to $L_{\rm EM}$ in the kinetic mixing case, $\dot{E}_{\rm rad}$ is suppressed (instead of enhanced) compared with $\dot{E}_{\rm GW}$ by an additional factor of $M_c/M$. Therefore, the frequency derivative is altered both by the additional dissipation channel and the additional frequency shift:
\begin{align}
    \dot{f}^\gamma_{\rm GW}=\left(\frac{5\alpha^3}{8\pi M^2}+\frac{\alpha^5 A'^{-2}_c}{16\pi^2 M^2}\right)\left(\dot{E}_{\rm GW}+\dot{E}_{\rm rad}\right),
    \label{eq:fdot}
\end{align}
for the most unstable vector field configuration in the $\alpha\ll 1$ limit. Combining \eqref{eq:fdot} with the flux upper limit obtained above (and using \texttt{SuperRad} to compute $\dot{E}_{\rm GW}$), we can determine the impact of the leading self-interactions on $\dot{f}_{\rm GW}$ and its derivatives at the points of gravitational saturation, where $M_c=M_c^{\rm sat}$. In the following, we use this to find the critical coupling $\rho_r$ (defined in the main text), below which these frequency derivative modifications do not impact the GW search sensitivities.

\subsubsection{Critical field amplitudes}

\begin{figure}
\includegraphics[width=0.49\textwidth]{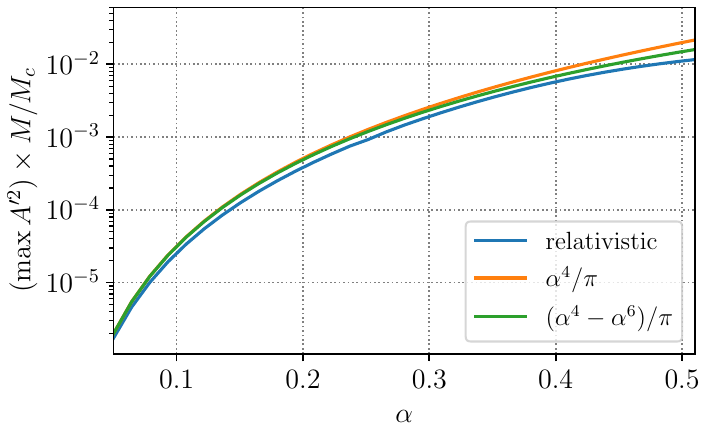}
\caption{The global maximum of the square of the vector field normalized by the cloud's mass $M_c$ on a Kerr background with $\chi=0.99$, obtained from test-field calculations \cite{Siemonsen2020} (labelled ``relativistic'') and leading non-relativistic estimates \cite{Baryakhtar2017} (labelled by their expressions). Here we focus entirely on the fastest growing superradiant field configuration. The maximum relative difference between the non-relativistic and the relativistic results is $<50\%$ at $\alpha\approx 0.5$.}
\label{fig:maxA}
\end{figure}

Here, we determine the critical couplings, $\rho_s$ and $\rho_r$, defined in the main text, which control the impact of the string formation and dark radiation on mapping null results obtained from GW searches for (Stueckelberg) vector clouds around CBC remnants to a dark Higgs-Abelian sector. Let us begin with the string production scenario. As outlined in the main text, string formation becomes efficient as soon as $\max A'^2$ approaches $A'^2_c=\lambda v^2/g^2$. In Ref.~\cite{East2022}, it was found that strings can begin to form when $\max A'^2$ is roughly an order of magnitude below $A'^2_c$. Additionally, $A'^2$ is gauge-dependent towards large amplitudes as the local U(1) symmetry is perturbatively restored, i.e., around $\max A'^2\lesssim A'^2_c$. Therefore, we introduce a conservative factor $\delta=10^2$ in the formation condition to take this into account and find $\max A'^2 \delta\lesssim A'^2_c$. The fastest growing superradiantly unstable vector field configuration at leading order in the expansion in $\alpha$ has the maximal amplitude $\max A'^2=(\alpha^4-\alpha^6)(M_c/M)/\pi$ \cite{Baryakhtar2017}. In Fig.~\ref{fig:maxA}, we compare this non-relativistic estimate against the more accurate fully relativistic test-field solutions obtained in Ref.~\cite{Siemonsen2020}. We find that the non-relativistic expression overestimates the maximal amplitude and that the largest relative difference across the different values of $\alpha$ is less than $50\%$. Hence, in the following, we conservatively use the non-relativistic expression for $\max A'^2 (M/M_c)$. All in all, if the maximal field amplitude at gravitational saturation is below this bound, i.e., 
\begin{align}
\delta\frac{M_c^{\rm sat}}{M}\frac{\alpha^4-\alpha^6}{\pi}\equiv A'^2_s \lesssim A'^2_c,
\end{align}
no strings are formed and the subsequent cloud evolution may be impacted by the presence of the Higgs only through the altered frequency evolution. The corresponding critical coupling
is $\rho_s=m_V^2A'^2_s$.
 
Turning to this modified GW frequency evolution, we determine the second critical coupling
$\rho_r$ by considering the maximal frequency time derivative that the search methods are still able to track. That is, just like in the kinetically mixed dark photon case, for successful tracking of the signal, the GW frequency (altered by the presence of the Higgs) may not wander out of the chosen frequency bin over a single coherent time segment. Consulting \eqref{eq:frequencycondition}, this condition is explicitly given by
\begin{align}
    \dot{f}^\gamma_{\rm GW}(0)-\frac{1}{2}\left[\dot{f}^\gamma_{\rm GW}(0)-\dot{f}^\gamma_{\rm GW}(T_{\rm coh})\right]=\dot{f}^{\gamma=0}_{\rm GW}(0),
    \label{eq:freqcondition2}
\end{align}
where $T_{\rm coh}=[2\dot{f}^{\gamma=0}_{\rm GW}(0)]^{-1/2}$. The frequency derivative after the first coherent time segment, $\dot{f}^\gamma_{\rm GW}(T_{\rm coh})$, is obtained by Taylor expanding $\dot{f}^\gamma_{\rm GW}$ around $t=0$ up to first order and determining $\ddot{f}^\gamma_{\rm GW}(0)$ from a derivative of \eqref{eq:fdot} and \eqref{eq:bosonicrad}. Equation~\eqref{eq:freqcondition2} is an implicit equation for $\gamma$, which defines the critical coupling $\rho_r=m_V^2/\gamma$ for which the GW emission after gravitational saturation is marginally detectable with a standard GW search assuming a Stueckelberg vector field. We find, in the parameter space relevant for this work, the correction to $\dot{f}^{\gamma=0}_{\rm GW}(0)$ originating from the shift \eqref{eq:deltaomgeaA} to be larger than those emerging from the dark radiation channel. In general, this coupling $\rho_r$ depends on $\alpha$ and the BH properties. As noted in the main text, constraints on the Stueckelberg vector field obtained from null observations apply to the dark Higgs-Abelian model for couplings $\lambda v^4\gtrsim\max(\rho_r,\rho_s)$.


%

\end{document}